\documentclass[twocolumn,prl,amsmath,amssymb,showpacs,superscriptaddress,floatfix]{revtex4}
\usepackage{color}
\usepackage{graphicx,epsfig}
\usepackage[hypertex]{hyperref}

\DeclareMathOperator{\Imm}{Im}

\DeclareMathOperator{\Rem}{Re}
\begin{document}
\author{N.\,M.\,Chtchelkatchev}
\affiliation{Argonne National Laboratory, Argonne, IL 60439,USA}
\affiliation{Department of Theoretical Physics, Moscow Institute
of Physics and Technology, 141700 Moscow, Russia}
\author{V.\,M.\,Vinokur}
\affiliation{Argonne National Laboratory, Argonne, IL 60439,USA}
\author{T.\,I.\,Baturina}
\affiliation{Argonne National Laboratory, Argonne, IL 60439,USA}
\affiliation{Institute of Semiconductor Physics, 13 Lavrentjev Ave., Novosibirsk, 630090 Russia}
\date{\today}

\title{Hierarchical energy relaxation in mesoscopic tunnel junctions: \\
Effect of nonequilibrium environment
on low-temperature transport}

\begin{abstract}
We develop a theory of far from the equilibrium transport in arrays of tunnel junctions.
We find that if the rate of the electron-electron interactions exceeds the rate of the
electron-phonon energy exchange, the energy relaxation
ensuring the charge transfer may occur sequentially.
In particular, cotunneling transport in arrays of junctions
is dominated by the relaxation via the intermediate bosonic \textit{environment},
the electron-hole excitations, rather than by the electron-phonon mechanism.
The current-voltage characteristics are highly sensitive to
the spectrum of the environmental modes and to the applied bias, which sets the
lower bound for the effective temperature.
We demonstrate that the energy gap in the electron-hole spectrum which opens
below some critical temperature $T^*$ due to long-range Coulomb interactions
gives rise to the suppression of the tunneling current.
\end{abstract}


\maketitle

Transport in mesoscopic tunnel junctions is ensured
by the energy exchange between the tunneling charge carriers and energy reservoirs:
since the electronic energy levels at the banks of the mesoscopic junctions are, in general,
different, the tunneling is impossible unless there is subsystem of excitations
capable of accommodation of this
energy difference~\cite{Nazarov1989,Devoret_basic,Averin1990,Girvin1990,Ingold1991,Ingold-Nazarov,Ingold1994}.
Intense studies of nano-structured and disordered systems including
Josephson junctions~\cite{Aprili2009},
mesoscopic superconductors~\cite{Timofeev2009},
patterned superconducting films~\cite{BaturinaSNS},
highly disordered superconducting
and semiconducting films~\cite{Shahar2005,Baturina2007,Shahar2009,other}
reveal a prime importance of the \textit{out-of-equilibrium} properties of an environment to
which the tunneling charge carriers relax the energy.
Notably, the relaxation processes can be mediated not only by phonons but by
the energy exchange with the
electromagnetic environment~\cite{Nazarov1989,Girvin1990,Averin1990,Ingold1991,Ingold-Nazarov,Ingold1994,Kopnin_book,Giazotto}
and with the electron-hole (e-h) pairs generated by the tunneling carriers~\cite{Lopatin1,Lopatin2}.
The energy relaxation in mesoscopic tunnel junctions in the case where
the energy exchange between the tunneling carriers and the electromagnetic
and/or electron-hole reservoir, $1/\tau_{\rm{env-e}}$,
is comparable to the rate of the energy loss to the phonon thermostat, $1/\tau_{\rm env\to bath}$,
was analyzed in~\cite{Kopnin2009}. In this Letter we
develop a general approach to the description of the strongly nonequilibrium
processes where $1/\tau_{\rm{env-e}}\gg 1/\tau_{\rm env\to bath}$
and show that the energy relaxation enabling the tunneling current occurs in two stages:
(i) The energy relaxation from the tunneling charges
to the intermediate bosonic modes (electromagnetic or electron-hole excitations)
which we hereafter call the \textit{environment}; and (ii)\,The energy transfer
from the environment to the phonon thermostat, to which we will be further referring as to
a \textit{bath}.

We demonstrate that the transport is controlled by the first stage and is thus
critically sensitive to the spectrum of the environmental modes.  At the same time,
the passing current drives the environment out of the equilibrium, and
the environment spectrum and effective temperature may become bias-dependent themselves.
We derive the coupled kinetics equations for the
charge carriers and out-of-equilibrium bosonic environment and apply our technique to
tunneling transport in large arrays of normal and superconducting junctions.

\begin{figure}[t]
\includegraphics[width=1\columnwidth]{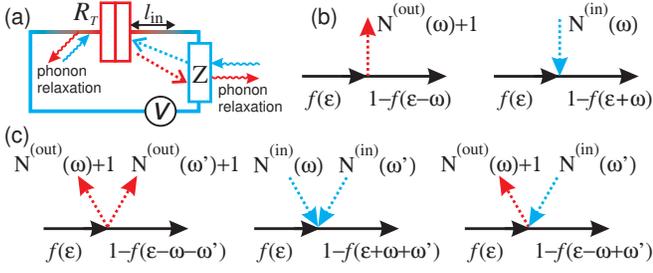}
\caption{(Color online.)
(a) The effective circuit for the tunnel junction subject to bias
$V$ and with the environment having the impedance $Z$.
(b)-(c) Diagrammatic expansion of $P^<$ to the first and the second orders in $\rho$ respectively.
The solid lines represent propagation of electrons,
the dashed lines denote the environment excitations.
The vertex with the two electron lines and one  dashed line
carries $G_{\scriptscriptstyle{\mathrm T}}\rho(\omega)/\omega$ factor, the
two dashed-lines vertex corresponds to $G_{\scriptscriptstyle{\mathrm T}}\rho(\omega)\rho(\omega')/(\omega\omega')$.}
\label{fig:e-ph_int}
\end{figure}

\textit{A single junction --}
First, we consider a tunnel junction between two bulk metallic electrodes biased
by the external voltage $V$, see Fig.\,\ref{fig:e-ph_int}a.
A general formula for the tunneling current reads:
          \begin{gather}\label{eq:current}
                  I=e\left(\overrightarrow{\Gamma} - \overleftarrow{\Gamma}\right)\, ,
           \end{gather}
where $\overrightarrow{\Gamma}$ ($\overleftarrow{\Gamma}$)
is the tunneling rate from the left (right) to the right (left), and, for a single junction,
   \begin{gather}\label{eq:S_begin_total}
        \overrightarrow{\Gamma}=\frac1{R_{\scriptscriptstyle {\mathrm T}}}
        \int_{\epsilon\epsilon'}f_\epsilon^{(1)}
        (1-f_{\epsilon'}^{(2)})P^<(\epsilon-\epsilon')\, ,
   \end{gather}
where $f^{(1,2)}$ are the electronic distribution functions within the
electrodes, $P^<(\epsilon)$
is the probability for the charge carrier to lose the energy $E$ to the environment, and
$R_{\scriptscriptstyle{\mathrm T}}$ is the bare tunnel resistance.
The backward scattering
rate, $\overleftarrow{\Gamma}\propto\int_{\epsilon\epsilon'}f_\epsilon^{(2)}
(1-f_{\epsilon'}^{(1)})P^<(\epsilon-\epsilon')$. If  an intermediate environment is
absent and the relaxation is provided by the phonon bath, then
$P^<(\epsilon)=\delta(\epsilon)$ and  Eq.~\eqref{eq:S_begin_total}
reproduces the conventional Ohm law.
The quasi-equilibrium situation
where the distribution functions of the environmental modes $N_{\omega}$
are Bose distributions parameterized
by the equilibrium temperature was discussed in~\cite{Grabert_Devoret}.
In a general, far from the equilibrium case, we find:
\begin{gather}\label{eq:S_begin_total1}
   P^{<}(E)=\int_{-\infty}^\infty dt \exp[J(t)+iEt]\,,
  \\ \label{eq:J} J(t)=2\int_{0}^\infty \frac{d\omega}{\omega}\rho(\omega) F(\omega)\,,
    \\\label{eq:S_begin_total1}
    F(\omega)=\left[N_\omega e^{i\omega t}+(1+N_\omega)e^{-i\omega t}-B_\omega\right]\,.
\end{gather}
The nonequilibrium distribution function
$N_{\omega}$ is defined by the kinetic equation
with the scattering
integral describing the energy exchange between environmental modes and tunneling
electrons.  Terms proportional to the
$N_\omega$ and $1+N_\omega$
correspond to the absorbed and emitted environmental excitations respectively.
The combination $B_\omega=1+2N_\omega$
is the kernel of the time-independent contribution to $J$ describing the elastic
interaction of the tunneling electron with the environmental modes
and having the structure of the Debye-Waller factor.
In an equilibrium, $N_\omega$ reduces to the
Bose-function and the functional $P^<$ recovers the result by
Ref.~\cite{Grabert_Devoret}.
The spectral probability of the electron--(electromagnetic) environment interaction is
$\rho(\omega)=
\Rem[Z_{\mathrm t}(\omega)]/R_{\scriptscriptstyle{\mathrm Q}}$,
where $Z_{\mathrm t}$ is the total circuit impedance
and $R_{\scriptscriptstyle{\mathrm Q}}$ is the  quantum resistance~\cite{Ingold1991}.
Proceeding analogously to
Ref.~\cite{Grabert_PRL}, one finds the spectral probability corresponding
to the electron--environment interaction within the each electrode as
$\rho_{\mathrm i}(\omega)
=2\Imm\int_{\mathbf q}\tilde U_{\mathrm i}/(D_{\mathrm i} q^2-i\omega)^{2}$, $i=1,2$,
 and that for the interaction across the junction,
$\rho_{12}(\omega)=-2\Imm\int_{\mathbf q}\tilde U_{12}/
[(D_1q^2-i\omega)(D_2q^2-i\omega)]$, where $D_{1(2)}$
are diffusion coefficients within respective electrodes, and $\tilde U_{1(2)}$
are the dynamically screened Coulomb interactions
within (across) the electrodes.
The form of $\rho(\omega)$ depends on the structure of the environmental
excitations spectrum and, thus, on the external bias.
The latter is especially important
in the array of highly transparent junctions where $\rho(\omega)$ is different for elastic and
inelastic processes~\cite{Nazarov1989,SVB}].
In particular, for the e-h environment with the constant $\tilde U$,
one should cut off the integral
at $q=\sqrt{T_{\rm eff}/D}$, when calculating $\rho(0)$, where $T_{\rm eff}$
is the (bias dependent) effective temperature of the environment which we determine below.
This allows us to formulate a recipe: if in an equilibrium $\rho=\rho(\omega,T)$ then in an
out-of-equilibrium state $\rho=\rho(\omega,T_{\rm eff})$.

To close the set of formulas (\ref{eq:current})-(\ref{eq:S_begin_total1})
one has to add the kinetic equations (KE) for the boson distribution functions $N_{\omega}$.
To derive these KE we use a semi-phenomenological kinetic approach of~\cite{Landauvshiz_10}
and express the current of Eq.\,(\ref{eq:current}) through the
electronic distribution function
as $I=\int_{\epsilon_1}\nu_1[d  f^{(1)}_{\epsilon_1}/dt]$.
Here $df^{(1)}_{\epsilon_1}/{dt}=I_{\rm col}$,
where $I_{\rm col}$ is the collision integral
describing the evolution of the electronic distribution function due to
energy and/or momentum transfer processes.
Expanding further $P^<$ with respect to $\rho$ we obtain, in the zero order in $N_{\omega}$,
the collision integral in a form
$I_{\rm col}^{(0)}=
-\int W_{12}[f^{(1)}_{\epsilon_1}(1-f^{(2)}_{\epsilon_2})-
f^{(2)}_{\epsilon_2}(1-f^{(1)}_{\epsilon_1})]
\delta(\epsilon_1-\epsilon_2)\nu_2 d\epsilon_2$,
where $W_{12}=1/\nu_1\nu_2R_{\scriptscriptstyle {\mathrm T}}$
is proportional to the
bare probability for an electron to be transmitted from one lead to the other.
In the first order
\begin{gather*}
\begin{split}
&\frac{df^{(1)}_{\epsilon_1}}{dt}=-\int d\omega\nu_\omega \nu_2d\epsilon_2\left(\frac{\rho}{\omega\nu_\omega}\right)W_{12}\times\\
&\biggl\{
\delta(\epsilon_{12}-\omega)[f^{(1)}_{\epsilon_1}(N_\omega+1)
(1-f^{(2)}_{\epsilon_2})-(1-f^{(1)}_{\epsilon_1})N_\omega f^{(2)}_{\epsilon_2}]+
\\
&\delta(\epsilon_{12}+\omega)[f^{(1)}_{\epsilon_1}N_\omega
(1-f^{(2)}_{\epsilon_2})-(1-f^{(1)}_{\epsilon_1})
(N_\omega+1) f^{(2)}_{\epsilon_2}]\biggr\},
\end{split}
\end{gather*}
where $\epsilon_{12}=\epsilon_1-\epsilon_2$ and $\nu_\omega$
is the density of environmental states \cite{env_DoS}.
The structure of
$I_{\rm col}^{(1)}$ is identical to that of the electron-phonon scattering integral in
metals \cite{Landauvshiz_10}, where $N_{\omega}$ would stand for
the phonon distribution functions.
The quantity $\rho/\omega\nu_\omega$ is proportional to the
probability of the electron-environment scattering.

The collision integral dual to $I_{\rm col}^{(1)}$
and describing the evolution of $N_{\omega}$ is derived analogously, and the
resulting kinetic equation is:
\begin{gather}\label{eq:N_e-env}
\left(\frac{dN_\omega}{dt}\right)_{\rm e-env}=
-\frac{A\rho(\omega)}{\nu_\omega R_{\scriptscriptstyle{\mathrm T}}}
\left[N_\omega (1+n_\omega)-(1+N_\omega)n_\omega\right],
\end{gather}
where $A$ is the numerical factor of order of unity, $n_\omega$
is the electron-hole pairs distribution function.
The scattering integral in Eq.\eqref{eq:N_e-env} is also identical by its structure to the
phonon-electron scattering integral in metals \cite{Landauvshiz_10}.
For the electron-hole environment ($i=1,2$ label the electrode in which the pair is located),
one has
$n_\omega^{(i)}=
(1/\omega)\int_\epsilon f^{(i)}_{\epsilon_+}(1-f^{(i)}_{\epsilon_-})$;
this agrees with the results of Ref.~\cite{Kamenev-Andreev} where the nonequilibrium
boson distribution function is equivalent to our $1+2n_\omega$.
If electrons and holes belong to different electrodes,
$n_\omega=
(2\omega)^{-1}\int_\epsilon f^{(i)}_{\epsilon_+}\sigma^x_{ij}(1-f^{(j)}_{\epsilon_-})$,
 $\hat\sigma^x$ being the Pauli-matrix.
From \eqref{eq:N_e-env} one estimates the rate of the energy exchange
between the environment and
the tunneling electrons as:
$1/\tau_{\rm{env-e}}=\rho(\omega)/
\nu_\omega R_{\scriptscriptstyle{\mathrm T}}$.
Now one has to compare
$1/\tau_{\rm{env-e}}$ with
the rate of the interaction of the environment modes with the (phonon) bath,
$1/\tau_{\rm env\to bath}(\omega)$.
For the electron-hole environment, $1/\tau_{\rm env\to bath}(\omega)$
is determined from Eq.\eqref{eq:N_e-env}
to which the electron-phonon scattering integral is added.
If $\tau_{\rm env\to bath}\gg \tau_{\rm env-e}$,  the two-stage relaxation takes place
and the characteristic energy transfer from tunneling current is
$\omega\sim \max\{T_{\mathrm e},V\}$, where
$T_{\mathrm e}$ is the electronic temperature in the leads.
The electromagnetic environment mediates the two stage relaxation in the case where Ohmic
losses occur in a LC superconducting line and are small~\cite{env_DoS}.
To take a typical example, in aluminum mesoscopic samples $\tau_{\rm env-e}=10^{-8}$\,sec
and $\tau_{\rm env\to bath}=10^{-6}$\,sec~\cite{Giazotto},
so the conditions for the two-stage relaxation are realized.
In this case the distribution functions, $N_\omega$,
of the environmental modes become nonequilibrium and are
determined from the condition that the collision integral
of the environmental modes with the e-h pairs accompanying
the current flow becomes zero, then Eq.\eqref{eq:N_e-env}
yields $N_\omega\cong n_\omega^{(12)}$.
If $T_{\mathrm e}\ll V$, then $N_\omega$ can
be approximated by the Bose-function with some effective
temperature $T_{\rm eff}$ at $\omega<V=T_{\rm eff}$ and $N_\omega=0$ at $\omega>T_{\rm eff}$
(the emission of the excitations with  the energy larger than $V$ is forbidden), and
\begin{gather}\label{eq:T_eff}
T_{\rm eff}\equiv\lim_{\omega\to 0} \omega N_\omega=0.5 V\coth(V/2T)\,.
\end{gather}
Thus the system with the environment well isolated
from the bath cannot be cooled below $T_{\rm eff}$.

Equations \eqref{eq:current}-\eqref{eq:N_e-env} give the full
description of the kinetics of the tunneling junction in a
nonequilibrium environment.  To derive the
$I$-$V$ characteristics we find $N_\omega\cong n_\omega^{(12)}$
and plug it into Eqs.\eqref{eq:current}-\eqref{eq:J}.
Introducing the parameters $g=\rho(0)$ and $\Lambda$,
the characteristic frequency of the
$\rho(\omega)$ decay [for the
Ohmic model~\cite{Grabert_Devoret}, $\rho=g^{-1}/\{1+(\omega/\Lambda)^2\}$ and
$\Lambda/g$ is of the order of the charging energy of the tunnel junction], we find:
\begin{gather}\label{eq:I_g}
I\sim \frac V {R_{\scriptscriptstyle{\mathrm T}}}\ln \frac \Lambda V\,;
\end{gather}
in the interval $T\ll V\ll \Lambda$, where $T_{\rm eff}\simeq V$.
Note that $I(V)$ given by
Eq.\eqref{eq:I_g} differs from the power law dependence
obtained in~\cite{Grabert_Devoret}
for $T_{\mathrm e}=T_{\rm eff}=0$.
This shows that tuning  the environment
one can manipulate by the tunnel junction $I(V)$ (the gating effect).
At high voltages, $V\gg\Lambda$, one finds
\begin{gather}
    I(V)\simeq(V-\Delta_\infty)/{R_{\scriptscriptstyle{\mathrm T}}}\,,
  \end{gather}
$\Delta_\infty=iJ'(0)=2\int_0^\infty d\omega\rho_{\omega}[1+N_\omega^{\rm (out)}-N_{\omega}^{\rm (in)}]\simeq\Delta_\infty^{(0)}\ln(\Lambda/\min\{T_e,T_{\rm env}\})$,
where $\Delta_\infty^{(0)}=\Delta_\infty[N^{\rm (out)}= N^{\rm (in)}]\sim \Lambda/g$,
since at $V\gg\Lambda$, $N^{\rm (out)}_{\omega}\simeq \Lambda/\omega\gg N^{\rm (in)}_{\omega}$.

\begin{figure}[t]
  \includegraphics[width=0.9\columnwidth]{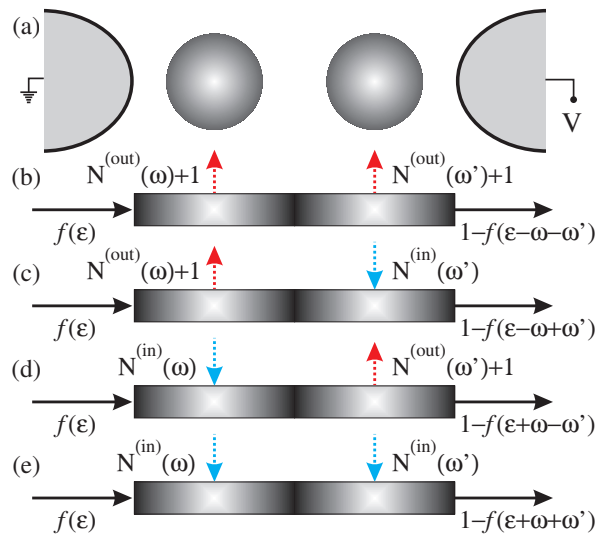}
  \caption{(Color online.) (a) The single electron two-islands circuit.
(b)-(e) Diagrams describing the forward inelastic cotunneling rate.
The  ``up" arrows stand for the e-h pairs excited during the cotunneling
and the ``down" arrows correspond to the recombination of the e-h pairs.
The vertices shown by boxes are proportional to the probability
of an elemental e-h pair excitation, $\rho(\omega)/\omega$.
}
\label{fig:cotunneling}
\end{figure}

\textit{Arrays of tunnel junctions--}
Extending Eq.\,(\ref{eq:S_begin_total}) onto an array comprised of $N$ junctions one finds
\begin{gather}\label{eq:cotunneling_rate}
    \overrightarrow{\Gamma}=\left(\prod_{i=1}^N\frac{R_K}{4\pi^2 R_{i}}\right)\,S^2\,\int d\epsilon d\epsilon' f_1(\epsilon)[1-f_2(\epsilon')] P(\epsilon-\epsilon'),
\end{gather}
where
\begin{multline}
    P(E)= \int_{-\infty}^\infty dt \exp(iEt)
    \left\{\int_{0}^\infty d\omega\frac{\rho(\omega)}\omega \times\right.
    \\ \left.\prod_{j\leq N-1}
    \left[N_{\omega,j}^{\rm (in)} e^{i\omega t}+(1+N_{\omega,j}^{\rm (out)})e^{-i\omega t}\right]\right\}\, .\label{eq:P(E)-gen}
\end{multline}
Here $S=E_{\mathrm c}^{-(N-1)}{N^N}/{(N-1)!}$, and $E_{\mathrm c}=e^2/2C$
is the Coulomb charging energy of a single junction ($C$
is a single junction capacitance) and for the Cooper pair transport $e\to 2e$.
 Eqs.~\eqref{eq:cotunneling_rate},\eqref{eq:P(E)-gen} were
derived in a first order in tunneling Hamiltonian.
Shown in Fig.\ref{fig:cotunneling} is a diagrammatic representation
of Eq.\eqref{eq:P(E)-gen} for $N=3$.

A generalization of the results obtained for a single junction including
the structure of the collision integral and the concept of the effective
temperature Eq.\,\eqref{eq:T_eff}, onto large arrays is straightforward.
As long as temperatures are not extremely low~\cite{Lopatin1}, the
charge transfer in large arrays is dominated by
the inelastic cotunneling and the two-stage energy relaxation.
The tunneling carriers generate e-h pairs~\cite{Lopatin1,Lopatin2} serving
as an environment exchanging the energy with the tunneling current and then
slowly losing it to the bath.
It is instructive to consider a two-dimensional array of superconducting tunnel junctions.
On the distances $L<\lambda=\sqrt{C/C_0}$, where $C_0$ is the capacitance of a
single junction to the ground, the Coulomb interaction between charges is logarithmic.
If the size of an array does not exceed $\lambda$, the e-h
plasma comprising the environment experiences the charge Berezinskii-Kosterlitz-Thouless
(BKT) transition~\cite{BKT,Fazio} at
$T=T_{\scriptscriptstyle{\mathrm{BKT}}}\simeq E_{\mathrm c}$~\cite{Fazio}.
This implies that at $T\simeq E_{\mathrm c}$ the energy gap $T^*$ opens
in the spectrum of unbound electrons and holes and, as a result,
$\rho(\omega)$ vanishes in the interval $0<\omega<T^*$. One than sees
from Eq.\,(\ref{eq:cotunneling_rate}) that
opening the gap suppresses
both Cooper pairs- and normal quasiparticle currents in the
superconducting tunneling array at $T<E_{\mathrm c}$.
Analyzing contribution from higher orders into
cotunneling process, one finds
that the current suppression holds in all orders.
This picture applies to the films
close to superconductor-insulator transition (SIT)~\cite{Baturina2007}.
Indeed, near the SIT the dielectric constant $\varepsilon$
of the film diverges~\cite{Dubrov}
and on the distances $L<\varepsilon d$, where $d$ is the film thickness,
the 2D e-h environment experiences the BKT transition.
Thus opening the gap in the electron-hole spectrum due to long range Coulomb effects
and the resulting suppression of the
tunneling current offers a microscopic mechanism for the
insulator-to-superinsulator transition~\cite{FVB,Nature}.

Two notion of the two-stage relaxation is a key to resolving the controversy of
the variable range hopping (VRH) conductivity in both doped
semiconductors~\cite{Khondaker1999,Shlimak1999} and
disordered superconducting films~\cite{Baturina2007}:
the observed universal pre-exponential factor indicates that the energy relaxation
is due to electron-electron (e-e) rather than the electron-phonon interactions.
On the other hand, according to~\cite{Gornyi2005,BAA2006} e-e relaxation
cannot ensure a finite conductivity below the
so called many-body localization temperature~\cite{BAA2006}.
The sequential relaxation of hopping electrons via
the e-h environment, which further transfers energy to the phonon bath
implies that the prefactor in hopping conductivity is indeed proportional to $e^2/\hbar$.

In conclusion, we have developed a quantitative description
of the highly nonequilibrium tunneling transport
in arrays of tunnel junctions in the limit $1/\tau_{\rm{env-e}}\gg 1/\tau_{\rm env\to bath}$
and demonstrated that the low-temperature relaxation ensuring the
tunneling current occurs via an intermediate
electromagnetic and/or e-h pairs environment.
We argued that the onset of the gap in the spectrum of
environmental excitations  suppresses tunneling current.
In particular, the gap due to Coulomb interactions in superconducting arrays
can offer a microscopic mechanism for the insulator-superinsulator transition.

We are grateful to R. Fazio, A. Shytov, A. Gurevich,
I. Burmistrov and Ya. Rodionov for useful discussions.
This work was supported by the U.S. Department of Energy Office of Science
under the Contract No. DE-AC02-06CH11357,
by the Programs of the Russian Academy of Sciences,
and by the Russian Foundation for Basic Research
(Grant Nos. 09-02-01205 and 09-02-12206).


\begin{thebibliography}{99}
\bibitem{env_DoS}
In an LC-line environment the excitations spectrum is linear,
$\omega=v k$, where $v=c/\sqrt {LC}$, so $\nu_\omega=1/v$.
\end{thebibliography}
\end{document}